\documentclass[english,aps,article,twocolumn]{revtex4}
\usepackage{times}
\usepackage[T1]{fontenc}
\usepackage[latin1]{inputenc}
\usepackage{graphicx}

\AtBeginDocument{
  
}

\makeatletter
\usepackage{babel}
\makeatother
\begin{document}

\title{Resonant x-ray diffraction study of the magnetoresistant perovskite
$\textrm{Pr}_{0.6}\textrm{Ca}_{0.4}\textrm{MnO}_{3}$.}

\author{S. Grenier,$^{1,2}$ J. P. Hill,$^{2}$ Doon Gibbs,$^{2}$ K. J.
Thomas,$^{2}$ M. v. Zimmermann,$^{3}$ C. S. Nelson,$^{4}$ V. Kiryukhin,$^{1}$Y.
Tokura,$^{5}$ Y. Tomioka,$^{5}$ D. Casa,$^{6}$ T. Gog,$^{6}$ C.
Venkataraman.$^{6}$ }

\affiliation{$^{1}$ Department of Physics and Astronomy, Rutgers University,
Piscataway, New Jersey 08854\\
$^{2}$ Department of Physics, Brookhaven National Laboratory, Upton,
New York 11973\\
$^{3}$Hamburger Synchrotronstrahlungslabor (HASYLAB) at Deutsches
Elektronen-Synchrotron (DESY), Notkestr. 85, 22603 Hamburg, Germany.\\
$^{4}$NRL-SRC, NSLS, Brookhaven National Laboratory, Upton, New York
11973\\
$^{5}$Joint Research Center for Atom Technology (JRCAT), Tsukuba
305-0046, Japan\\
$^{6}$ CMC-CAT, Advanced Photon Source, Argonne National Laboratory,
Argonne, Illinois 60439\\
}

\begin{abstract}
We report a resonant x-ray diffraction study of the magnetoresistant
perovskite $\textrm{Pr}_{0.6}\textrm{Ca}_{0.4}\textrm{MnO}_{3}$.
We discuss the spectra measured above and below the
semiconductor-insulator transition temperature with aid of a detailed
formal analysis of the energy and polarization dependences of the
structure factors and \emph{ab initio} calculations of the spectra.
In the low temperature insulating phase, we find that inequivalent Mn
atoms order in a CE-type pattern and that the crystallographic
structure of La$_{0.5}$Ca$_{0.5}$MnO$_3$, (Radaelli \emph{et al.},
Phys. Rev. B 55, 3015 (1997)) can also describe this system in fine
details. Instead, the alternative structure proposed for the so-called
Zener polaron model (Daoud-Aladine \emph{et al.}, Phys. Rev. Lett. 89,
097205 (2002)) is ruled out by crystallographic and spectroscopic
evidences. Our analysis \emph{supports} a model involving orbital
ordering. However, we confirm that there is no direct evidence of
charge disproportionation in the Mn \emph{K}-edge resonant spectra.
Therefore, we consider a CE-type model in which there are two Mn
sublattices, each with partial $e_{g}$ occupancy.  One sublattice
consists of Mn atoms with the $3x^{2}-r^{2}$ or $3y^{2}-r^{2}$
orbitals partially occupied in a alternating pattern, the other
sublattice with the $x^2-y^2$ orbital partially occupied.  
\end{abstract}
\maketitle

\section{Introduction.}

The interplay among the various electronic degrees of freedom,
including those of spin, charge, lattice and orbital degeneracy lies
at the heart of the wide variety of phenomena observed in strongly
correlated electron materials. These include unusual transport
properties observed in colossal magnetoresistive (CMR) manganites and
high-temperature superconductivity in cuprates. Particularly
noteworthy examples of this interplay occur in the perovskite
manganites $\textrm{RE}_{x}\textrm{AE}_{1-x}\textrm{MnO}_{3}$ (where
$\textrm{RE}$ is a trivalent rare earth and $\textrm{AE}$ a divalent
alkaline earth) for which the Mn atoms have a partially filled, high
spin, \emph{3d} band. These materials exhibit rich phase diagrams in
which the balance between the various degrees of freedom may be
altered by a variety of methods, including hole doping, cationic size
mismatch, temperature, pressure, magnetic field and electromagnetic
radiation \cite{PhysicsOfManganites,SalamonRMP}.

One of the most interesting ground states that occurs in these phase
diagrams arises in the vicinity of half doping (\emph{x} = 0.5). This
is a phase which has been believed to exhibit charge, orbital and
magnetic order. It is exhibited in a number of compounds, including
amongst others $\textrm{Pr}_{x}\textrm{Ca}_{1-x}\textrm{MnO}_{3}$
\cite{Jirak85}, $\textrm{La}_{x}\textrm{Ca}_{1-x}\textrm{MnO}_{3}$
\cite{Wollan55,Radaelli97}, and
$\textrm{Nd}_{0.5}\textrm{Sr}_{0.5}\textrm{MnO}_{3}$ \cite{Nakamura99}
as well as some other layered manganites such as
$\textrm{La}_{1.5}\textrm{Sr}_{0.5}\textrm{MnO}_{4}$
\cite{Sternlieb95}.  Further, closely related phases have been
observed in cobaltates, \emph{e.g.}
$\textrm{La}_{1.5}\textrm{Sr}_{0.5}\textrm{CoO}_{4}$
\cite{Zaliznyak01} and nickelates, \emph{e.g.}
$\textrm{La}_{1.5}\textrm{Sr}_{0.5}\textrm{NiO}_{4}$
\cite{Kajimoto03}. In addition to its ubiquity, it is interesting as
an example of the balance among the various degrees of freedom and
because in Manganites it exhibits the CMR effect: It is an
antiferromagnetic insulating phase, but application of a magnetic
field melts the charge and orbital order (COO), driving the formation
of a ferromagnetic metallic state and thus causing a dramatic decrease
in the resistivity. Recent theories suggest that this phenomena is
driven by a competition between charge ordered phases - such as the
CE-type - and ferromagnetic metallic regions in a
phase-separation-type picture \cite{dagotto01}.

The search for a microscopic picture of the ground-state in half-doped
manganites remains a very active field. In the 1950's, Goodenough
described the ordering as comprising a checkerboard pattern of
Mn$^{3+}$ and Mn$^{4+}$ sites (charge order) \cite{Goodenough55}. In
this picture, the Mn$^{3+}$ sites have an extra $e_{g}$ electron that
occupies a $3z^{2}-r^{2}$-type orbital and these orientationally align
in a cooperative manner to form an anti-ferro-type pattern within the
plane (orbital order). On the basis of the exchange pathways set up by
this order, a complex magnetic ordering occurs which may be thought of
as zig-zag chains of ferromagnetically aligned spins which are coupled
antiferromagnetically with their neighbors (CE-type AF order). A
schematic of this ordering is shown in Fig. \ref{fig Goodenough
model}.

\begin{figure}
\includegraphics[width=0.75\columnwidth]{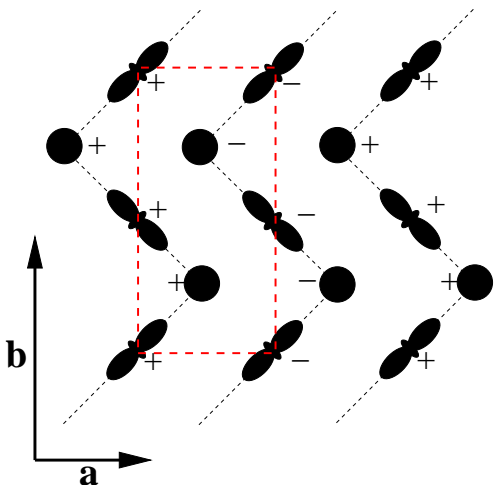}
\caption{\label{fig Goodenough model}Schematic of the CE-type charge,
orbital and magnetic ordering as described by Goodenough in 1955
\cite{Goodenough55}.  The elongated figure-eights represent the
occupied $e_{g}$ $(3x^{2}-r^{2})$-type orbitals on the Mn$^{3+}$
sites, the closed circles represent the Mn$^{4+}$ sites. The signs +
and - indicate the relative orientations of the spin. The
ferromagnetic zigzag chains are indicated by the dotted lines, the
rectangle indicates the low-temperature unit cell.}
\end{figure}

While debate continues as to the origin of the stability of this
phase, this original picture has survived relatively unchallenged to
the present day, garnering significant theoretical and experimental
support.  Experimentally, strong evidence includes the various
structural studies (both x-ray and neutron) which reveal the presence
of inequivalent Mn sites, one of which sits in a distorted octahedron
consistent with $3z^{2}-r^{2}$ occupancy, the other in an undistorted
octahedron (see e.g. \cite{Radaelli97}). Further, neutron refinements
of magnetic moments find two different moments on the two sites, with
$\mu$(3+)/$\mu$(4+) \textasciitilde{} 1.1 - 1.2 (for instance in
$\textrm{Nd$_{0.5}$Sr$_{0.5}$MnO$_{3}$}$ \cite{Kawano97},
$\textrm{La}_{0.5}\textrm{Ca}_{0.5}\textrm{MnO}_{3}$
\cite{Radaelli97}, $\textrm{La$_{0.5}$Sr$_{1.5}$MnO$_{4}$}$
\cite{Sternlieb95} and $\textrm{Pr$_{0.6}$Ca$_{0.4}$MnO$_{3}$}$
\cite{Jirak85}). In addition, the observed magnetic structure is
consistent with expectations for this charge and orbitally ordered
structure based on the so-called Goodenough-Kanamori-Anderson rules
\cite{Kanamori}. Further consistency is found with recent resonant
x-ray scattering results which identified short-range orbital
correlations as the origin of the observed short-range magnetic
correlations on the Mn$^{3+}$ sublattice
\cite{Zimmerman01a,Radaelli97}.  Finally, transport, optical and NMR
data have all been interpreted in terms of this picture
\cite{PhysicsOfManganites}.

Concerning the theory, several groups have argued as to the dominant
mechanism leading to the stability of the CE-type phase, but the basic
picture has not been questioned
\cite{Anisimov97,vdBrink99,Oles01,Mahadevan01,Popovic02}.  However,
\emph{ab initio} calculations predict a non-integer mixed valence
accompanying an orbitally ordered CE-type phase in half-doped
manganites.  For $\textrm{Pr}_{0.5}\textrm{Ca}_{0.5}\textrm{MnO}_{3}$,
Anisimov \emph{et al.} \cite{Anisimov97} showed that two different Mn
$e_{\textrm{g}}$ configurations exist which have almost the same
charge density with different orbital configurations. The COO is
predicted to be of the checkerboard type, one site having a
$3x^{2}-r^{2}$-type symmetry, the other a $x^{2}-y^{2}$-type
symmetry. Using the local-spin-density approximation and including the
intra-\emph{d}-shell Coulomb interaction, the calculation was
performed without the \emph{a priori} constraint of the CE-type
pattern and the concomittant Jahn-Teller distortions.  A similar
description based on density functional theory using the gradient
approximation has also recently been presented for half-doped
La$_{0.5}$Ca$_{0.5}$MnO$_{3}$ \cite{Popovic02}. In addition
J. v.d. Brink \emph{et al.} \cite{vdBrink99} proposed that partial
charge ordering occurs due to the strong Coulomb repulsion on one site
with partly occupied $x^{2}-y^{2}$ and $3z^{2}-r^{2}$ orbitals,
whereas the adjacent site is occupied either by the $3x^{2}-r^{2}$ or
$3y^{2}-r^{2}$ orbitals.

Despite this body of evidence and the consistency of the COO picture,
the nature, the pattern and even the existence of the Mn valence
organization have been recently questioned. In particular, a complete
valence separation (\emph{i.e.} $\textrm{Mn}^{3+}/\textrm{Mn}^{4+}$)
appears to be inconsistent with high-resolution X-Ray Absorption Near
Edge Structure (XANES) spectra \cite{Garcia01sg}. Manganites with and
without COO, show similar XANES spectra which cannot be made up of the
sum of the spectra from the parent compounds that have integer valence
of 3+ or 4+. These studies suggest that no, or only a small charge
disproportionation, either on the manganese atoms or at a molecular
scale, can be supported \cite{Garcia01sg}. More explicit is the recent
crystallographic structure refinement of Daoud-Aladine \emph{et al.}
\cite{Daoud-Aladine02} which was performed without the \emph{a priori}
constraints of the mixed-valence pattern: the resulting structure is
inconsistent with the CE-type model, it exhibits no significant charge
disproportionation and serves as a basis for introducing a new model
based on so-called Zener polarons. Finally, the resonant x-ray
scattering data have been criticized on the grounds that they are
mostly sensitive to the position of the oxygen atoms and that it is
possible that a purely structural distortion could result in the same
scattering patterns \cite{Garcia01}.  Thus, despite 50 years of
experimental and theoretical effort, there remain some very basic
questions that remain to be answered in the half-doped manganites.

In this paper, we seek to address this issue by performing resonant
x-ray diffraction (RXD) studies of
$\textrm{Pr}_{0.6}\textrm{Ca}_{0.4}\textrm{MnO}_{3}$, which is
believed to exhibit CE-type charge and orbital order. As discussed
below, this technique, when accompanied by detailed analysis, is
extremely sensitive to the environment of the resonant ion (in this
case Mn) and thus the details of the electronic ordering. In
particular we analyse the RXD spectra from both above and below the
structural phase transition. Our main result from the low temperature
studies is that inequivalent Mn atoms do in fact order in the CE-type
pattern. We argue that on one of the sites (``3+'') there is indeed
$3x^{2}-r^{2}/3y^{2}-r^{2}$-type ordering of the \emph{3d} $e_{g}$
orbitals. However, we find no evidence for a chemical shift of the
\emph{1s} levels and interpret this as an absence of significant
charge disproportionation. Therefore on the basis of our data, we
suggest that the partial occupancy on the ``4+'' site is in the
$x^{2}-y^{2}$-type orbital. Finally, all our experimental observations
rule out the crystallographic structure upon which the Zener polaron
model was based, they are however consistent with the XANES studies.

In section \ref{sec:Material-and-Experimental}, we provide details of
the sample and the resonant x-ray experiments. In section
\ref{sec:Experimental-Results.}, the spectra measured at room
temperature ($T>T{}_{COO}$) and in the low temperature phase
($T{}_{COO}>T>T_{N}$) are presented. The results are discussed in
section \ref{sec:Discussion.} and analysed with the aid of \emph{ab
initio} calculations. We summarize our results in section
\ref{sec:Summary.}.

\section{\label{sec:Material-and-Experimental}Material and Experimental Methods.}

\begin{figure}
\includegraphics[width=0.75\columnwidth]{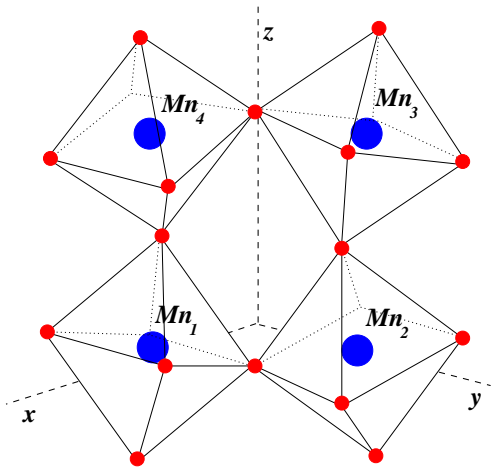}
\caption{\label{fig unit cell}High temperature unit cell in the
\emph{Pbnm} perovskite structure of Pr$_{0.6}$Ca$_{0.4}$MnO$_{3}$. The
oxygen octahedra are all equivalent; Mn$_{2}$ , Mn$_{3}$ and Mn$_{4}$
are related to Mn$_{1}$ by the \emph{b}, \emph{n} and \emph{m} mirrors
respectively. Pr/Ca atoms (not shown) lie between the octahedra
layers.}
\end{figure}

At room temperature
$\textrm{Pr}_{0.6}\textrm{Ca}_{0.4}\textrm{MnO}_{3}$ has the $Pbnm$
structure \cite{Jirak85} (see Fig. \ref{fig unit cell}).  At low
temperatures, Jir\'ak \emph{et al.} \cite{Jirak85} find that the
ground state is a CE-type antiferromagnet ($T_{N}\approx170$ K),
exhibiting charge and orbital order ($T_{COO}\approx232$ K).  The
magnetic structure consists of ferromagnetic Mn zig-zag chains coupled
antiferromagnetically in the (\textbf{a},\textbf{b}) plane and stacked
ferromagnetically \emph{}along the \textbf{c} direction.

The COO phase in $\textrm{Pr}_{0.6}\textrm{Ca}_{0.4}\textrm{MnO}_{3}$
is evidenced by a sudden decrease in the magnetic susceptibility, an
increase in the resistivity at $T_{COO}>T_{N}$
\cite{Tomioka96,Okimoto99} and by the appearance of superstructure
Bragg reflections which indicate a doubling of the unit cell. These
reflections also disappear when the magnetic field drives the compound
into the metallic state\cite{Zimmerman01a}.  X-ray studies of the
superlattice reflections above the transition suggest that charge
ordering drives the orbital ordering \cite{Zimmerman01a}.

We have chosen the $\textrm{Pr}_{1-x}\textrm{Ca}_{x}\textrm{MnO}_{3}$
system for a number of reasons: (i) CMR is observed in a commensurate
COO phase which is stabilized for a range of doping $0.3<x<0.7$
\cite{Tomioka96,Simon02}.  (ii) The similar size of the Ca and Pr
cations reduces strain effects.  (iii) A further study of the
electronic and magnetic phase transitions in this material is
advantageous because of the significant difference between the
magnetic ordering and COO transition temperatures.

The sample was prepared at the Joint Research Center for Atom
Technology in Japan; the growth and the basic transport properties
have been described in detail elsewhere \cite{Tomioka96}. The $(010)$
surface was polished with a 1$\mu$m grit and the mosaic of the sample
was $0.25^{\circ}$ FWHM as measured at the (020) reflection.

The experiments were performed utilizing resonant x-ray diffraction
which involve measuring the intensity of a reflection as a function of
the incident photon energy
\cite{Materlik,Murakami98,Benfatto99,Paolasini99,vettier01}.  By
tuning the incident energy to the Mn absorption \emph{K}-edge,
\emph{1s} electrons are promoted to an intermediate unoccupied
\emph{p-}state and then decay back to the \emph{1s}. Therefore one
probes the unoccupied density of \emph{p-}states projected onto the Mn
atoms as a function of energy. Because these intermediate states
reflect any structural anisotropy, the scattering factors of the Mn
atoms are no longer scalars but become tensors.  In addition, the
coherence of the resonant process implies that diffraction can occur
and we can probe the long-range-ordered correlations of the local
electronic configuration: This technique combines spectroscopic
information with that of a scattering experiment.

The Mn \emph{4p} states are sensitive to the surrounding structure
because their spatial distribution extends out to, and beyond, the Mn
nearest neighbors. In order to interpret the RXD spectra measured at
the \emph{K}-edge, we therefore make the assumption that the
particular structural distortions of the Mn surroundings reflect the
highest occupied \emph{3d} orbitals. We note that the characteristic
time for the resonant process is about $10^{-16}$ s while that of the
lattice vibrations is $10^{-12}$ s and therefore RXD provides a
snap-shot of the distortions surrounding the Mn atoms.

Interpreting these particular resonant spectra is still
complex. According to previous works in the litterature, it is not
possible to draw conclusions about the surrounding structure, the
charge or orbital ordering without detailed, quantitative modelling
and analysis of the various contributions to the resonant
scattering. Below, we provide a description of the resonant structure
factor within the dipolar approximation, both above and below the
phase transition temperature. These calculations provide an
understanding of the characteristics of the resonant signals which
allow us to infer the characteristics of the electronic
configuration. In particular, we are interested in determing the
symmetries of the highest occupied orbitals.

\begin{figure}
\includegraphics[width=0.75\columnwidth]{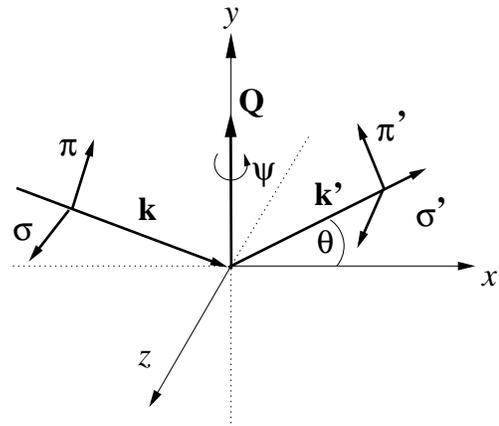}
\caption{\label{fig diffraction geometry} The diffraction geometry. An
azimuthal scan consists of rotating the sample by an angle $\psi$
around the diffraction vector \textbf{Q}=\textbf{k'-k}. The vectors
$\sigma$ and $\pi$ are the basis for the polarization vector of the
photon.}
\end{figure}

The x-ray experiments were performed at the CMC-CAT undulator beamline
9IDB at the Advanced Photon Source (Argonne National Laboratory) and
at beamline X22C at the National Synchrotron Light Source (Brookhaven
National Laboratory). Beamline 9IDB possesses a double-crystal Si
(111) monochromator with an energy resolution of $\Delta
E/E\approx2.10^{-4}$.  Beamline X22C has a double Ge (111)
monochromator with a resolution $\Delta E/E\approx5.10^{-4}$. We have
focused on the incident energy dependence of the diffracted intensity,
as it is tuned through the Pr-\emph{L}$_{II}$ and Mn\emph{-K}
absorption edges which in this oxide are at 6444 eV and 6552 eV
respectively, where we have defined the position of the edge by the
maximum of the first derivative of the absorption spectrum. In this
paper we focus on the vicinity of the Mn \emph{K-}edge energy. For
some of the data collected, the scattering was resolved into the
respective $\sigma-\sigma^\prime$ and $\sigma-\pi^\prime$ polarization
channels, where we adopt the standard notation that $\sigma$ ($\pi$)
denotes the polarization perpendicular (parallel) to the scattering
plane (see Fig. \ref{fig diffraction geometry}). The polarization
analysis was performed by utilizing a Cu (220) analyzer crystal for
which $2\theta_{Bragg}\approx96^{\circ}$ at the Mn \emph{K}-edge.
This discrepancy from the ideal 90 degrees leads to an expected
leakage of about $\cos^{2}(96)\approx1.1\%$ for the projection of one
polarization component into the other one. By measuring the fully
$\sigma-\sigma^\prime$ polarized Bragg reflection (020) in the
$\sigma-\pi^\prime$ channel analyzer we measured the leakage to be
about 1.5\%. High \textbf{Q} resolution measurements were performed
with a Ge(111) analyzer crystal. In addition to the diffraction
experiment, complementary XANES measurements were performed on the
same sample. These two techniques probe the same resonant scattering
factors, though the XANES measurement lacks the site-selectivity of
resonant diffraction, it provides directly the average of the
imaginary part of the resonant scattering factor. These latter
measurements were carried out at room temperature at beamline X11A
(National Synchrotron Light Source, Brookhaven National Laboratory).
This beamline has a double crystal Si (111) monochromator. The 6.539
keV \emph{K}-edge of a Mn foil was used to calibrate the energy at all
beamlines.

\section{\label{sec:Experimental-Results.}Experimental Results.}

\subsection{High temperature phase ($T>T_{COO}$).}

\begin{figure}
\includegraphics[width=0.75\columnwidth]{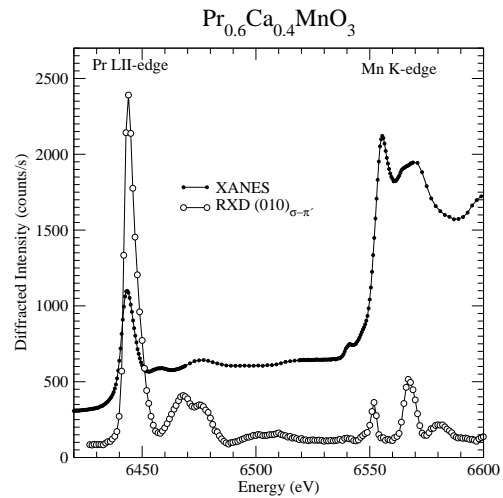}
\caption{\label{fig 010 RT} (open circles) Resonant X-ray Diffraction
(RXD) of the $\left(010\right)_{\sigma-\pi^\prime}$ \emph{Pbnm}-forbidden
peak measured in the $\sigma-\pi^\prime$ channel at 280K
($T>\textrm{T}_{COO}$) through the Pr \emph{L$_{II}$}-edge (6444 eV)
and the Mn \emph{K}-edge (6552 eV). The incident polarization is
directed along the \textbf{a} axis, \emph{i.e.} $\psi=90^{\circ}$. The
room temperature X-ray Absorption Near Edge Structure (XANES) spectra
(closed circles) is also shown over the same energy interval.}
\end{figure}

Resonant x-ray diffraction (RXD) spectra were collected in the high
temperature phase in order to provide a baseline to compare to the low
temperature ordered phase. Fig. \ref{fig 010 RT} shows the RXD
spectrum of the forbidden (010) reflection in the $\sigma-\pi^\prime$
channel taken over a wide energy range at 280 K, i.e. well above the
phase transition. It has no $\sigma-\sigma^\prime$ component. The
XANES spectrum at room temperature is also shown on the same energy
scale.

A number of features are seen in the RXD spectrum, prominantly at the
Pr $L_{II}$ and Mn \emph{K}-edges. At the Pr $L_{II}$ -edge one
observes three peaks, the main peak coinciding with the maximum of the
absorption spectra. At the Mn \emph{K}-edge, the spectrum shows three
peaks each with a Gaussian line-shape with half-widths that increase
as a function of the energy. In contrast to the Pr $L_{II}$-edge, the
first peak at $E_{i}=$ 6552 eV coincides with the maximum of the first
derivative of the absorption spectrum. As we will discuss below, the
origin of this signal is due to the loss of the exact octahedral
symmetry around the Mn atoms as a result of the tilt ordering. The
$(010)_{\sigma-\pi^\prime}$ reflection has long range order, that is,
its width in reciprocal space, as measured at the Mn \emph{K}-edge
with a Ge (111) analyzer, is similar to that of the (020) Bragg
reflection.  Thus, this scattering at (010) is distinct from that of
the pre-transitional fluctuations just above $T_{COO}$ observed in the
$\sigma-\sigma^\prime$ channel, which exhibit temperature-dependent
short-range order \cite{Zimmerman01a}.  Rather, this scattering
represents the average long-ranged ordered component of the
high-temperature structure (Templeton scattering \cite{Templeton80}).

The XANES measurements were performed by measuring the total
fluorescence yield with the beam along the $[010]$ direction and with
the incident polarization vector along both \textbf{a} and
\textbf{c}. In fact, this rotation of 90 degrees around the incident
direction produced no measurable change in the spectral features.  It
seems likely that the footprint of the beam is sufficient to overlap
different a-, b- and c-domains in this twinned sample such that the
resulting XANES spectrum is an average over these domains and thus
independent of the nominal polarization direction.

\subsection{Low temperature phase ($T<T_{COO}$).}

\begin{figure}
\includegraphics[width=0.75\columnwidth]{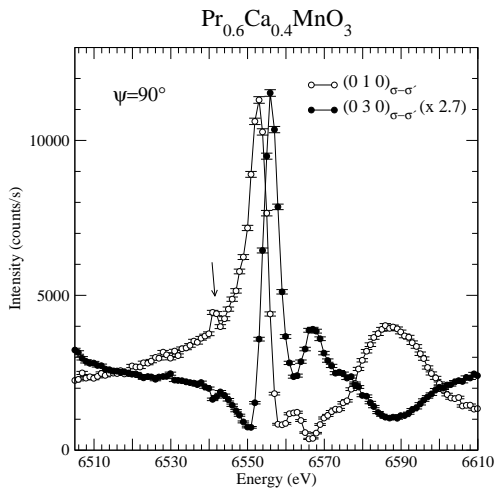}
\caption{\label{fig: (010) and (030) DANES spectra}RXD spectra of the
$\left(010\right)_{\sigma-\sigma^{\prime}}$ and
$\left(030\right)_{\sigma-\sigma^{\prime}}$ reflections at 100 K. The
3 eV difference between the two maxima is due to the different
crystallographic structure factors \cite{Zimmerman01a}. The small
feature at 6542 eV (arrow) is attributed to the pre-edge transition.
This feature was not seen in the previous data \cite{Zimmerman01a}
because of the lower resolution of that data set. The incident
polarization is along the \textbf{a} axis, \emph{i.e.}
$\psi=90^{\circ}$.}
\end{figure}

In the low temperature phase, v. Zimmermann \emph{et al.}
\cite{Zimmerman01a} reported and discussed the spectra of the (010),
(030), $(0\frac{1}{2}0)$ and $(0\frac{5}{2}0)$ reflections, together
with their polarization dependence. In Fig. \ref{fig: (010) and (030)
DANES spectra}, we report new data for the energy dependence of the
$(010)_{\sigma-\sigma^\prime}$ and $(030)_{\sigma-\sigma^\prime}$. These data were
taken with a higher energy resolution, that is 1.5 eV compared to the
earlier 5 eV resolution \cite{Zimmerman01a}. The overall shape is the
same as the earlier data with no change in the energy widths of the
observed features, indicating that they were not resolution limited in
the earlier data set. That is, the observed widths are determined by
the finite lifetime of the excited electron-hole pair and by band
structure effects.

There is however a small difference between these data and the earlier
scans, namely the feature observed around 6542 eV in both spectra in
Fig. \ref{fig: (010) and (030) DANES spectra}. We attribute this to
the pre-edge transitions (i.e. $1s\rightarrow3d$) - possibly
dipole-allowed from the breaking of inversion symmetry at the Mn
site. Such transitions are expected to be relatively sharp and thus
would have been smeared out in the earlier, lower resolution data.

\begin{figure}
\includegraphics[width=0.75\columnwidth]{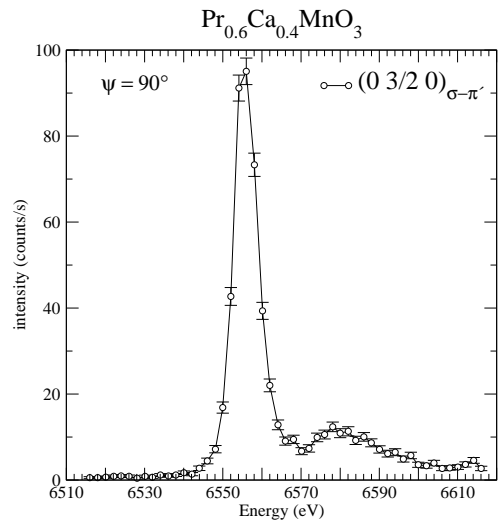}
\caption{\label{fig orbital spectra}RXD spectrum of the
$(0\frac{3}{2}0)_{\sigma-\pi^{\prime}}$ reflection at 10 K reproduced
from \cite{Zimmerman01a}. The energy dependence is strongly
reminescent of the Jahn-Teller compound $\textrm{LaMnO}_{3}$ which has
an orbital ordering of the highest \emph{3d} orbital occupied
(Fig. \ref{fig orbital spectra lmo}). The energy resolution for these
data was about 5 eV.}
\end{figure}

For completness we reproduce the $(0\frac{3}{2}0)_{\sigma-\pi^\prime}$
RXD of v. Zimmermann \emph{et al.} \cite{Zimmerman01a} in
Fig. \ref{fig orbital spectra} and show unpublished data of the
$(010)_{\sigma-\pi^\prime}$ measured on $\textrm{LaMnO$_{3}$}$ in the
orbital ordered phase in Fig. \ref{fig orbital spectra lmo}. Note that
the orbital order in LaMnO$_3$ has a propagation vector equals to
(010), whereas in Pr$_{0.6}$Ca$_{0.4}$MnO$_3$ it is equals to (020).

The spectrum of the $(010)_{\sigma-\pi^\prime}$ reflection of
Pr$_{0.6}$Ca$_{0.4}$MnO$_3$ was also measured at low
temperature. Unfortunately, at these temperatures, there is
significant scattering in the $\sigma-\sigma^\prime$ channel at this
wave-vector and care must be taken that leakage from this channel is
not falsely ascribed to $\sigma-\pi^\prime$ scattering. The severity
of this problem is illustrated in Fig. \ref{fig leakage}, which shows
that the 1.5\% leakage of the present analyzer is sufficient to
account for almost all of the apparent $\sigma-\pi^\prime$ scattering
at \emph{T}=180 K in the first ($E_{i}=6552$ eV) and third
($E_{i}=6580$ eV) features.

Note that no absorption correction has been made in these spectra.  In
addition, the spectra were checked to be free of spurious multiple
scattering by rotating the sample around the diffraction vector.

\section{\label{sec:Discussion.}Discussion.}

\subsection{\label{sec:Discussion HT}High temperature phase ($T>T_{COO}$).}

\begin{figure}
\includegraphics[width=0.75\columnwidth]{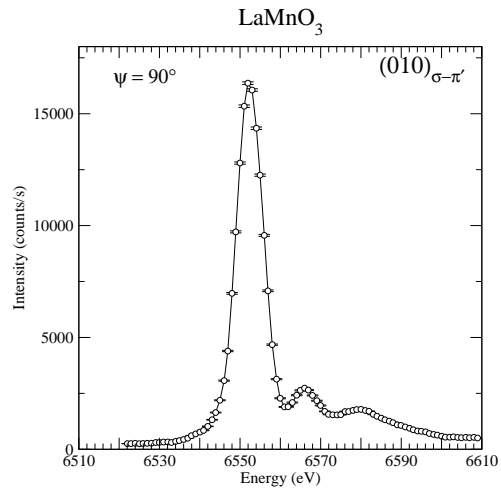}
\caption{\label{fig orbital spectra lmo} RXD spectrum of the
$(010)_{\sigma-\pi^{\prime}}$ of $\textrm{LaMnO$_{3}$}$. The
strong resonance arises from the Jahn-Teller distortion of the oxygen
octahedra due to the orbital ordering \cite{Murakami98}.  The energy
resolution for these data was about 5 eV.}
\end{figure}

\begin{figure}
\includegraphics[width=0.75\columnwidth]{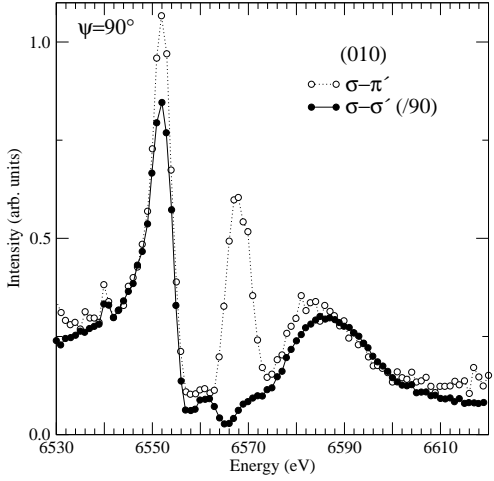}
\caption{\label{fig leakage}Measurement of the two scattering channels
with the Cu(220) analyzer in the vicinity of the Mn \emph{K}-edge at
the (010) reflection at 180 K. The figure illustrates the
contamination of the $\sigma-\pi^{\prime}$ channel measurement by the
$\sigma-\sigma^{\prime}$ channel. The intensity from the
$\sigma-\sigma^{\prime}$ channel has been scaled to show that a
leakage about 1\% leads to a significant contamination of the
$\sigma-\pi^{\prime}$ channel in the ordered phase. This complicates
the analysis of the low temperature phase (see text). However, the
second resonance at about 6570 eV is almost uncontaminated.}
\end{figure}

Above the phase transition temperature $T_{COO}=232$ K, the
crystallographic structure is described by the \emph{Pbnm} space group
with one Mn atom sitting at four equivalent sites \cite{Jirak85}. The
structure is orthorhombic, pseudo-cubic, and the lattice paramaters
are $a=5.4315$ \AA, $b=5.446$ \AA and $c=7.6481$ \AA (throughout this
paper all crystallographic notations will refer to the \emph{Pbnm}
unit cell, even at low temperatures where the space group symmetry is
actually lowered). Fig. \ref{fig unit cell} shows the unit cell with
the four Mn sites that are related by the symmetries of the space
group \emph{Pbnm.}  The Mn are situated in oxygen octahedra. The
structure shows the $\textrm{GdFeO}_{3}$-type distortion, that is the
octahedra are actually tilted from the \textbf{c}-axis in the
\textbf{a} direction \cite{Woodward97,Zimmerman01b}. They are
compressed along the \textbf{c} axis (Mn-O$_{z}$ = 1.9544 \AA) and
expanded in the (\textbf{ab}) plane (Mn-O$_{xy}$ = 1.9738 \AA,
Mn-O$_{x\overline{y}}$ = 1.9714 \AA).

In the following we derive several resonant structure factors of the
Mn atoms. Similar approaches have been taken previously by Murakami
\emph{et al.} \cite{Murakami98}, Takahashi \emph{et al.}
\cite{Takahashi99} and Garc\'ia \emph{et al.} \cite{Garcia01}. These
expressions will serve as a comparison with those derived later for
the low temperature phase. In order to quantify the resonant x-ray
cross-section of this distorted structure one has to take into account
the resulting anisotropy of the crystal field around the Mn sites. The
scattering amplitude is then described as a tensor of rank two within
the dipolar approximation (E1) \cite{Materlik}.  In the following
\emph{x, y} and \emph{z} are defined along the crystallographic axes
(see Fig. \ref{fig unit cell}). By assigning to one of the Mn atoms,
Mn$_{1}$, the most general dipolar tensor, $f_{1}$ , and then applying
the mirror symmetries of the \emph{Pbnm} space group, as represented
by the matrices $M_{x},\, M_{y}\,\textrm{and }M_{z}$, one can generate
the scattering tensors for the four equivalent Mn atoms:

\begin{eqnarray}
f_{1} & = & \left(
\begin{array}{ccc} f_{xx} & f_{xy} & f_{xz}\\ 
f_{xy} & f_{yy} & f_{yz}\\ f_{xz} & f_{yz} & f_{zz}\end{array}
\right),\nonumber\\ 
f_{2} & = & M_{x}f_{1}M_{x}=\left(\begin{array}{ccc} f_{xx} &
-f_{xy} & -f_{xz}\\ -f_{xy} & f_{yy} & f_{yz}\\ -f_{xz} & f_{yz} &
f_{zz}\end{array}\right),\nonumber \\
f_{3} & = & M_{y}f_{1}M_{y}=\left(\begin{array}{ccc} f_{xx} & -f_{xy} &
f_{xz}\\ -f_{xy} & f_{yy} & -f_{yz}\\ f_{xz} & -f_{yz} &
f_{zz}\end{array}\right), \nonumber \\
f_{4} & = & M_{z}f_{1}M_{z}=\left(\begin{array}{ccc} f_{xx} & f_{xy} &
-f_{xz}\\ f_{xy} & f_{yy} & -f_{yz}\\ -f_{xz} & -f_{yz} &
f_{zz}\end{array}\right).
\label{eq: resonant matrices}
\end{eqnarray}

When the incident energy is tuned to the Mn absorption edge, the
scattering power of each of these crystallographically equivalent
atoms may be represented by these matrices. One sees that as a result
of the symmetry operators, they are not the same. The scattering then
becomes polarization dependent and the standard crystallographic
reflection conditions are altered \cite{Dmitrienko83}. The
off-diagonal terms of the scattering tensor give rise to these effects
and are non-zero at the absorption edge because the intermediate
electronic states are anisotropic. As a result, the Mn sites give an
anomalous contribution to the $Pbnm$ forbidden reflections such as
$(h00)$, $(0k0)$ or $(0k\ell)$, whereas the contributions of O, Ca and
Pr, which have isotropic scattering factors, cancel exactly. We
consider below the total structure factors for the Mn atoms at several
forbidden reflections:

\begin{eqnarray}
F^{Mn}(h00) & = & F^{Mn}(0k0) \nonumber \\
& = & f_{1}-f_{2}-f_{3}+f_{4} = 4f_{xy}\left(\begin{array}{ccc}
0 & 1 & 0\\
1 & 0 & 0\\
0 & 0 & 0\end{array}\right),\nonumber \\
F^{Mn}(00\ell) & = & f_{1}+f_{2}-f_{3}-f_{4} = 4f_{yz}\left(\begin{array}{ccc}
0 & 0 & 0\\
0 & 0 & 1\\
0 & 1 & 0\end{array}\right),\nonumber \\
F^{Mn}(0k\ell) & = & f_{1}-f_{2}+f_{3}-f_{4} = 4f_{xz}\left(\begin{array}{ccc}
0 & 0 & 1\\
0 & 0 & 0\\
1 & 0 & 0\end{array}\right),\nonumber\\
\label{eq:resonant str. factors}
\end{eqnarray}
with $h,\, k\,\textrm{and}\,\ell$ being odd. Different energy dependences
of the resonant spectra are expected for these reflections since each
probes different components of the Mn resonant scattering tensor.

We next use these tensors to calculate the polarization and azimuthal
dependences of the intensity within the two common experimental
geometries $\sigma-\sigma^{\prime}$ and $\sigma-\pi^{\prime}$ (see
Fig. \ref{fig diffraction geometry}).  We define the azimuthal angle
$\psi$ as the angle between the incident polarization vector $\sigma$
and the \textbf{c}-axis, for the $(h00)$ and $(0k0)$ reflections, and
the angle between $\sigma$ and the \textbf{a}-axis for the $(00\ell)$
and $(0k\ell)$ reflections.  An azimuthal scan of one particular
reflection corresponds to measuring the intensity on rotating the
sample an angle $\psi$ in the plane perpendicular to the diffraction
vector. With these definitions, the polarization vectors are:

\begin{eqnarray}
\sigma(h00) & = & (0,\,-\sin\psi,\,\cos\psi),\nonumber \\
\pi^{\prime}(h00) & = &
(\cos\theta_{B},\,-\sin\theta_{B}\cos\psi,\,-\sin\theta_{B}\sin\psi),\nonumber
\\ \sigma(0k0) & = & (\sin\psi,\,0,\,\cos\psi),\nonumber \\
\pi^{\prime}(0k0) & = &
(-\sin\theta_{B}\cos\psi,\,\cos\theta_{B},\,\sin\theta_{B}\sin\psi),\nonumber
\\ \sigma(00\ell) & = & (\cos\psi,\,\sin\psi,\,0),\nonumber \\
\pi^{\prime}(00\ell) & = &
(\sin\psi\sin\theta_{B},\,-\cos\psi\sin\theta_{B},\,\cos\theta_{B}),\nonumber
\\ \sigma(0k\ell) & = &
(\cos\psi,\,\sin\psi\cos\alpha,\,-\sin\psi\sin\alpha),\nonumber \\
\pi^{\prime}(0k\ell) & = &
(\sin\theta\sin\psi,\,\nonumber\\
&&\cos\theta_{B}\sin\alpha-\sin\theta_{B}\cos\psi\cos\alpha,\nonumber\\
&&\sin\theta_{B}\cos\psi\sin\alpha+\cos\theta_{B}\cos\alpha) \nonumber \\
\end{eqnarray}
where $\theta_{B}$ is the Bragg angle and
$\alpha=\textrm{atan}\frac{c}{b}$.  Then, for example, the scattering
from the $(0k\ell)$ reflection for a $\sigma-\pi^{\prime}$ diffraction
geometry is given by
$I_{\sigma-\pi^{\prime}}(0k\ell)=|\sigma(0k\ell)F(0k\ell)\pi^{\prime}(0k\ell)|^{2}$.
The intensities corresponding to the above structure factors are:

\begin{eqnarray}
\label{eq:resonant intensities}
I_{\sigma-\sigma^{\prime}}(h00) & = & I_{\sigma-\sigma^{\prime}}(0k0) = 0,\\
I_{\sigma-\pi^{\prime}}(h00) & = &I_{\sigma-\pi^{\prime}}(0k0) = |4f_{xy}\cos\theta_{B}\sin\psi|^{2},\\
I_{\sigma-\sigma^{\prime}}(00\ell) & = & 0, 
I_{\sigma-\pi^{\prime}}(00\ell) = |4f_{yz}\cos\theta_{B}\sin\psi|^{2},\\
I_{\sigma-\sigma^{\prime}}(0k\ell) & = & |4f_{xz}\sin2\psi\sin\alpha|^{2},\\
I_{\sigma-\pi^{\prime}}(0k\ell) & = & |4f_{xz}|^2 \times \nonumber\\
& & |(\sin\theta_{B}\sin\alpha\cos2\psi +
\cos\theta_{B}\cos\alpha\cos\psi)|^{2}. \nonumber\\
\end{eqnarray}

These calculations show that $I_{\sigma-\pi^{\prime}}(0k0)$ and
$I_{\sigma-\pi^{\prime}}(00\ell)$ have two-fold symmetry with respect
to $\psi$ and that the $\sigma-\sigma^{\prime}$ channel is non-zero
for the forbidden $(0k\ell)$
reflections. $I_{\sigma-\sigma^{\prime}}(0k\ell)$ and
$I_{\sigma-\pi^{\prime}}(0k\ell)$ are proportional to each other as a
function of the x-ray energy but differ in their azimuthal symmetries,
the $\sigma-\sigma^{\prime}$ channel being four-fold symmetric in
$\psi$ while the $\sigma-\pi^{\prime}$ channel has no particular
symmetry. All of these characteristics have been observed in
perovskite oxides of this space group
\cite{Murakami98,Nogushi00,Kubota??}.  An incident $\pi$ component can
also give a contribution to the $\pi^{\prime}$-polarized
scattering. However, considering the highly linear polarized
synchrotron source, the beamline optics and the vertical diffraction
geometry we used, the incident $\pi$ component is expected to be much
smaller than the incident $\sigma$ component and will not be
considered in the following (there also systematic extinctions, in
particular $F_{\pi-\pi^{\prime}}(0k0)=0$ at all $\psi$).

The particular azimuthal dependences arise due to a geometrical effect
- a result of the symmetries between the equivalent resonant atoms in
the \emph{Pbnm} space group. So, the azimuthal dependence \emph{per
  se} is independent of the anisotropy in the occupied and unoccupied
density of states. Rather a more detailed analysis is required before
a conclusion concerning orbital occupancies may be drawn. This
calculation of the resonant intensity is widely applicable since the
space group \emph{Pbnm} describes many other manganite, titanate and
vanadate perovskites \cite{Takahashi99,Takahashi01,Takahashi02}. Note
that this space group includes both compounds that are orbitally
ordered and some that are orbitally disordered, as in the present
case.

As a result of the tilting of the octahedra relative to the
crystallographic axes, non-zero off-diagonal elements are introduced
into the scattering tensor. That is, the off-diagonal terms come from
both the different lengths of the principal directions of the
octahedra (\emph{i.e.}  an asymmetry inside the octahedra), and from
the tilt of the octahedra off the crystallographic axes (asymmetry
outside the octahedra). One expects that the further the octahedra are
tilted from the polarization direction the more important the
off-diagonal terms become. Conversely, decreasing the degree of tilt
decreases the signal arising from the distortion of the octahedra: If
the principal axis of the octahedra were along the crystallographic
axes, then $f_{xy}=f_{xz}=f_{yz}=0$, and $F^{Mn}$ would become diagonal
and the signal would disappear. In a sense then, the intensity of the
whole resonant spectrum is modulated by the degree of octahedral tilt.

From Eq. \ref{eq:resonant intensities}, we see that the experimental
measurement of the $(010)_{\sigma-\pi^{\prime}}$ reflection at
$\psi=90^{\circ}$, measures $|f_{xy}|$. One notes that for this
reflection the resonance at the edge (6552 eV) has a lower intensity
than the second resonance at 6568 eV (Fig. \ref{fig 010
RT}). Interestingly, this is in contrast with the data for
$\textrm{LaMnO}_{3}$. The two compounds have a similar overall
structure in the same space group, but $\textrm{LaMnO}_{3}$ shows
orbital ordering \cite{Rodriguez98}. In $\textrm{LaMnO}_{3}$, the
first resonance for the same reflection $(010)_{\sigma-\pi^{\prime}}$
is much larger, and is understood as coming mainly from the in-plane
Jahn-Teller distortion that reveals the orbital ordering
\cite{Murakami98,Benfatto99,Elfimov99}.  For $\textrm{LaMnO}_{3}$ the
in-plane Mn-O distances are Mn-O$_{xy}$ = 1.907 \AA,
Mn-O$_{x\overline{y}}$ = 2.178 \AA \cite{Rodriguez98}.  For
$\textrm{Pr$_{0.6}$Ca$_{0.4}$MnO}_{3}$, the anisotropy in the plane at
room temperature is much smaller. The distances are: Mn-O$_{xy}$ =
1.971 \AA, Mn-O$_{x\overline{y}}$ = 1.974 \AA ~\cite{Jirak85}. This
lower anisotropy is revealed qualitatively by a much lower intensity
of the first resonance. Recently, Takahashi, Igarashi and Fulde
performed \emph{ab initio} calculations for $\textrm{LaMnO}_{3}$,
$\textrm{YTiO}_{3}$ and $\textrm{YVO}_{3}$ in the orbitally ordered
phases \cite{Takahashi99,Takahashi01,Takahashi02}.  These authors
completed calculations of the $\sigma-\pi^{\prime}$ channel without
including any ordering of the \emph{3d} orbitals, thereby emphasizing
the role of the structural distortions as an origin of the resonant
signal. From this, they infered that the first peak in the Mn spectra
arises mainly from the in-plane anisotropy, and that the second was
largely due to the tilt order. Thus, this interpretation of the
calculations seems consistent with the data presented in fig. \ref{fig
010 RT} and those of LaMnO$_{3}$, given the difference between the
in-plane anisotropy in the two cases. In addition, as discussed above,
we believe that the tilt also has an effect over the entire spectrum
(that is the influence of the tilt order is not confined to the second
peak).

In order to gain further insight into the electronic configuration of
the Mn in this doped manganite, we next present preliminary \emph{ab
initio} calculations of the resonant diffraction. In particular we
want to investigate whether the known structural distortions are
enough to give rise to the resonant signal observed in
$\textrm{Pr$_{0.6}$Ca$_{0.4}$MnO$_{3}$}$ above the charge and orbital
order temperature. Our calculations of the
$(010)_{\sigma-\pi^{\prime}}$ have been done in the framework of the
Full Multiple Scattering theory with the FDMNES code \cite{Joly01}.
Starting with the atomic positions and their associated electronic
densities, this code solves the Green's matrix for the intermediate
states to which the photoelectron is promoted. The structure used is
that of Jir\'ak \emph{et al.} \cite{Jirak85} but with all the RE sites
replaced by Ca - the ability to treat a random distribution of Pr and
Ca ions is beyond the scope of the present work. In addition, the
electronic configurations of the atoms are described as if they were
isolated, that is Mn is $3d^{5}4s^{2}$ and so on. The Hedin-Lundquist
exchange-correlation potential is used.

\begin{figure}
\includegraphics[width=0.75\columnwidth]{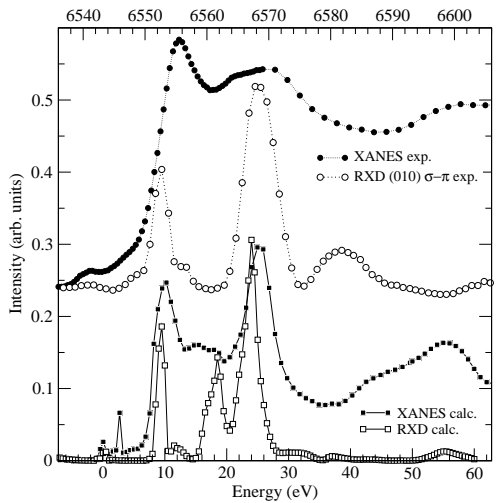}
\caption{\label{fig :DANES-and-XANES}RXD of the (010) reflection and XANES
from an \emph{ab initio} calculation (see text) compared to the experimental
spectra taken at room temperature. For the XANES spectra the polarization
is averaged both in the calculation and in the measurement.}
\end{figure}

The results of these calculations are shown in Fig. \ref{fig
:DANES-and-XANES}, which compares the XANES and the RXD measured at
room temperature with the calculations of each. The preliminary
results are quite satisfactory despite the simplistic approximations
used and the fact that no broadening for the core-hole lifetime or the
energy resolution have been applied.  The main disagreement is the
peak at 18 eV in the calculation of the RXD. These results suggest
that no detailed description of the Mn \emph{3d} orbitals is needed to
predict qualitatively the appearance of resonance peaks at high
temperature. All that is needed is the anisotropic structural
configuration.

\subsection{Low temperature phase ($T<T_{COO}$).}

Below $T_{COO}$, two sets of peaks appear at (i) Q+(010) and Q+(100),
and (ii) at Q+(0$\frac{1}{2}$0), where Q stands for the diffraction
vector of the Bragg reflections allowed in the \emph{Pbnm} space
group.  The latter superstructure peaks indicate a doubling of the
unit cell along the \textbf{b} direction. The intensity of these two
types of reflections shows the same temperature dependence; however,
they are differentiated by their different correlation lengths within
the low-temperature phase \cite{Zimmerman01b}. In the framework of the
COO picture, the Q+(0$\frac{1}{2}$0) reflections are due to
distortions induced by the orbital order. The orbital order forms in a
domain state with randomly distributed antiphase boundaries formed by
misoriented orbitals.  The domains are characterized by a correlation
length $\xi=320\pm10$ \AA~ \cite{Zimmerman01b,Radaelli97}. In contrast
the Q+(010) type reflections are resolution or near-resolution limited
with a correlation length $\xi>2000$ \AA.

A strong resonance effect is observed at the superstructure
reflections $(0\frac{3}{2}0)_{\sigma-\pi^{\prime}}$,
$(010)_{\sigma-\sigma^{\prime}}$ and $(030)_{\sigma-\sigma^{\prime}}$
when the polarization is in the (\textbf{ab}) plane
($\psi=90^{\circ}$) (Figs. \ref{fig: (010) and (030) DANES spectra}
and \ref{fig orbital spectra}). In particular, the resonant effect at
the Q+(010) reflection may be described qualitatively as a
{}``derivative effect'' \cite{Nakamura99,GrenierS01}, that is, the
lineshape in the energy scan has the form of the derivative of the
resonant factors.  This is a clear signal of the presence of one
element sitting at two \emph{different} crystallographic sites which
contribute to the structure factor with opposite phase.

In the literature, the appearance of the derivative effect has usually
been said to be a direct observation of charge ordering
\cite{Murakami98a,Nakamura99,GrenierS01,Staub02}.  The argument is
that, the decrease (increase) of the electronic density of the
resonant atom lowers (raises) the energy of the initial \emph{1s} core
level resulting in a tighter (weaker) binding energy. In this picture,
this shift applies directly to the x-ray resonant factors of the
Mn$^{4+}$ (Mn$^{3+}$) which shift to higher (lower) energy at the Mn
\emph{K}-edge (\emph{i.e.} for transitions from \emph{1s} to
\emph{p}-states). It is noteworthy that any \emph{1s} shift has an
isotropic effect on the spectra, whereas any shift of the \emph{4p}
will depend on the Mn-O distances and will therefore be
anisotropic. The combination of both shifts constitutes the chemical
shift.  For example, an isotropic shift occurs in the highly
anisotropic vanadium site in the charge ordered phase of
$\alpha^{\prime}-\textrm{NaV}_{2}\textrm{O}_{5}$ for which the
derivative effect has been observed for two perpendicular directions
of the incident polarization with the same energy shift
\cite{GrenierS01}. However, in the present compound the resonant
signal disappears when the polarization is along the \textbf{c} axis
($\psi=0^{\circ}$) \cite{Zimmerman01a}. Thus, the simplest picture of
an isotropic chemical shift of only the \emph{1s} core levels does not
apply here. This observation has also been made in other perovskites,
including manganites \cite{Nakamura99}. To understand the resonance
behavior in these materials therefore requires a more quantitative
study, which we develop in the following.

\subsubsection{Low temperature structure factor.}

\begin{figure}
\includegraphics[width=0.75\columnwidth]{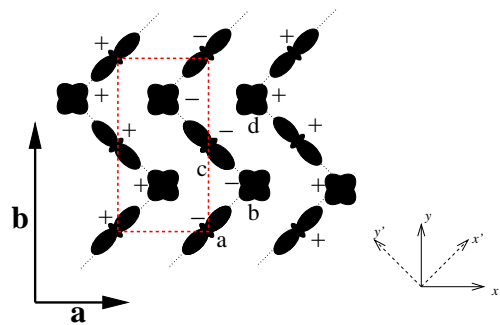}
\caption{\label{fig pattern}Schematic diagram of the modified CE-type structure
in the (\textbf{ab}) plane. The elongated Fig.-eights represent the
occupied $e_{g}$ $(3x^{\prime2}-r^{2})$-type orbitals, while the
clover shapes represent $e_{g}$ $(x^{\prime2}-y^{\prime2})$ orbitals.
The signs + and - indicate the spin direction in the magnetically
ordered phase. The letters denote the Mn for the model of the low
temperature phase only. The sites {}``a'' and {}``b'' correspond
respectively to the sites 1 and 2 in the high temperature phase. The
dashed line indicates the unit cell.}
\end{figure}

In the doubled low-temperature unit cell there are 8 Mn atoms. As
widely observed in the three-dimensional manganites
\cite{Radaelli97,Lees98,Daoud-Aladine02}, the planes are believed to
be equivalent along the \textbf{c}-axis, which leaves 4 independent Mn
atoms in the same plane (see schematic in Fig. \ref{fig pattern}). As
above we will keep the calculations within the dipolar
approximation. For example, the general dipole structure factor of the
Q+(010) peaks $\sigma-\sigma^{\prime}$ channel is:

\begin{eqnarray}
F_{\sigma-\sigma^{\prime}}^{Mn}(0k0) & = & 2\times[(f_{xx}^{(a)}-f_{xx}^{(b)}+f_{xx}^{(c)}-f_{xx}^{(d)})\sin^{2}\psi\nonumber \\
 &  & +(f_{zz}^{(a)}-f_{zz}^{(b)}+f_{zz}^{(c)}-f_{zz}^{(d)})\cos^{2}\psi\nonumber \\
 &  &
 +(f_{xz}^{(a)}-f_{xz}^{(b)}+f_{xz}^{(c)}-f_{xz}^{(d)})\sin2\psi],\nonumber\\
\label{eq: fac str}
\end{eqnarray}
where \emph{a}, \emph{b}, \emph{c} and \emph{d} label the 4 Mn sites
within the plane (Fig. \ref{fig pattern}). In contrast to the high
temperature phase, the $(0k0)$ reflections are no longer forbidden as
a result of very small displacements of the atoms. Therefore a term
almost constant in energy must be added to the total structure factor
to take into account the Thomson scattering and the resonant
corrections from other absorption edges from all the atoms, arising
from the fact that this scattering no longer precisely
cancels. However, since we are interested mainly in the resonant
contribution of the Mn atoms, and in any case these corrections are
small, we will ignore them in the following, except where explicitly
stated. As noted above, no resonance is measured at $\psi=0^{\circ}$,
so one infers from Eq. \ref{eq: fac str} that
$f_{zz}^{(a)}-f_{zz}^{(b)}+f_{zz}^{(c)}-f_{zz}^{(d)}$ is zero within
the limits of our experiment, the energy resolution of the experiment
being about 5 eV \cite{Zimmerman01a}. In contrast, the resonant effect
at $\psi=90^{\circ}$ indicates that
$f_{xx}^{(a)}-f_{xx}^{(b)}+f_{xx}^{(c)}-f_{xx}^{(d)}$ is non zero.

Similarly, the general dipole structure factor for the Q+$(0\frac{1}{2}0)$
peaks $\sigma-\pi^{\prime}$ channel is:

\begin{eqnarray}
\label{eq: str factor orb}
F_{\sigma-\pi^{\prime}}^{Mn}(0\frac{k}{2}0) & = &
 -(f_{xx}^{(a)}+if_{xx}^{(b)}-f_{xx}^{(c)}-if_{xx}^{(d)})\sin\theta_{B}\sin2\psi\nonumber
 \\ & &
 -(f_{xz}^{(a)}+if_{xz}^{(b)}-f_{xz}^{(c)}-if_{xz}^{(d)})\sin\theta_{B}\cos2\psi\nonumber
 \\ & &
 +2(f_{xy}^{(a)}+if_{xy}^{(b)}-f_{xy}^{(c)}-if_{xy}^{(d)})\cos\theta_{B}\sin\psi\nonumber
 \\ & &
 +2(f_{yz}^{(a)}+if_{yz}^{(b)}-f_{yz}^{(c)}-if_{yz}^{(d)})\cos\theta_{B}\cos\psi\nonumber
 \\ & &
 +(f_{zz}^{(a)}+if_{zz}^{(b)}-f_{zz}^{(c)}-if_{zz}^{(d)})\sin\theta_{B}\sin2\psi\nonumber
 \\
\end{eqnarray}
Using expressions such as equations \ref{eq: fac str} and \ref{eq: str factor orb}
we can proceed, constrained by the experimental data, to develop a
model for the low temperature electronic order.

\subsubsection{\label{sub:CE-type-model}Model of the CE-type electronic configuration.}

By considering various models of the COO phase, with particular
symmetries between the Mn atoms one can simplify Eq. \ref{eq: fac
str}. The simplest model of the COO phase, as considered by Garc\'ia
\emph{et al.}  \cite{Garcia01}, is a checkerboard model in which
Mn$_{b}$ and Mn$_{d}$ are identical and isotropic,
i.e. $f_{xx}^{(b)}=f_{yy}^{(b)}=f_{zz}^{(b)}=f_{xx}^{(d)}=f_{yy}^{(d)}=f_{zz}^{(d)}=f$
and the off-diagonal terms are zero. For the sites \emph{a} and
\emph{c}, they used an alternative notation that is
$f_{xx}^{(a)}=f_{yy}^{(c)}=\frac{f_{\parallel}+f_{\perp}}{2}$ and
$f_{xy}^{(a)}=f_{\parallel}-f_{\perp}$ where $\parallel$ ($\perp$)
indicates the direction parallel (perpendicular) to the stretching of
the $\textrm{e}_{\textrm{g}}$ orbital \cite{Garcia01}. This equivalent
description will be convenient in a case described later for
understanding the shape of the resonances. For our analysis, we make
the less restrictive ansatz that the electron density of the Mn$_{b}$
and Mn$_{d}$ sites have an square symmetry in-plane. At this point,
this is a starting assumption. However, as we shall see, it is both
consistent with our experimental data and recent theoretical
calculations. Such in-plane symmetry can include, for example, the
population of $\textrm{e}_{g}$ orbitals of the
$x^{\prime2}-y^{\prime2}$ or $3z^{\prime2}-r^{2}$ symmetry as
suggested by recent theoretical calculations \cite{vdBrink99} (in
labelling the orbitals we use the ($x^{\prime},\, y^{\prime},\,
z^{\prime}$) coordinate system aligned along the extension of the
e$_{g}$ orbitals, see Fig. \ref{fig pattern}, to preserve the more
familiar description of these orbitals). In such a model, we force
$f_{xx}^{(b)}=f_{xx}^{(d)}=f_{yy}^{(b)}=f_{yy}^{(d)}\neq
f_{zz}^{(b,d)}$.  Following the labels in Fig. \ref{fig pattern},
Mn$_{a}$ and Mn$_{c}$ have a $3x^{\prime2}-r^{2}$ and
$3y^{\prime2}-r^{2}$ geometry respectively.  They are related by a
$\frac{\pi}{2}$ rotation around the \textbf{c} axis (more accurately,
they are related by a $\frac{\pi}{2}$ rotation about the principal
axis of the octahedra which is tilted 10 degrees from the \textbf{c}
direction as a result of the tilt order), with the extension of the
highest occupied \emph{3d} orbital and a concomittant extension of the
Mn-O bonds along the $[110]$ and $[1\bar{1}0]$ directions. It follows
that $f_{xx}^{(c)}=f_{yy}^{(a)}$, $f_{xz}^{(c)}=f_{xz}^{(a)}$ and
$f_{xy}^{(c)}=-f_{xy}^{(a)}$. The resulting Mn structure factors in
this model are:

\begin{eqnarray}
\label{eq:LT str. factors}
F_{\sigma-\sigma^{\prime}}^{Mn}(0\frac{k}{2}0) & = &
2(f_{xx}^{(a)}-f_{yy}^{(a)})\sin^{2}\psi,\nonumber \\
F_{\sigma-\pi^{\prime}}^{Mn}(0\frac{k}{2}0) & = &
- (f_{xx}^{(a)}-f_{yy}^{(a)})\sin2\psi\sin\theta_{B} \nonumber \\ 
&& +  4f_{xy}^{(a)}\sin\psi\cos\theta_{B},\nonumber \\
F_{\sigma-\sigma^{\prime}}^{Mn}(0k0) & = &
2(f_{xx}^{(a)}+f_{yy}^{(a)}-2f_{xx}^{(b)})\sin^{2}\psi \nonumber \\
&& +  4(f_{zz}^{(a)}-f_{zz}^{(b)})\cos^{2}\psi \nonumber \\
&& +  4(f_{xz}^{(a)}-f_{xz}^{(b)})\sin2\psi],\nonumber \\
F_{\sigma-\pi^{\prime}}^{Mn}(0k0) & = &
4f_{xy}^{(b)}\sin\psi\cos\theta_{B} \nonumber \\
&& +  4(f_{yz}^{(a)}-f_{yz}^{(b)})\cos\psi\cos\theta_{B}\nonumber \\
&& -  2(f_{xz}^{(a)}-f_{xz}^{(b)})\cos2\psi\sin\theta_{B}\nonumber \\
&&
+  \sin2\psi\sin\theta_{B}\times \nonumber \\
&&
[2(f_{zz}^{(a)}-f_{zz}^{(b)})+2f_{xx}^{(b)}-(f_{xx}^{(a)}+f_{yy}^{(a)})]\nonumber
.\\
\end{eqnarray}

Note again that for simplicity these relations neglect the fact that
the Mn atoms are slightly displaced. This displacement will give rise
to small corrections to the expressions. However, for the present
purposes, we are concerned with which components will contribute to
the resonant scattering and the associated azimuthal dependence, for
which, these corrections are unimportant. Below, we write explicitly
these scattering factors for two limits of azimuthal geometries,
$\psi=90^{\circ}$ and $\psi=0^{\circ}$.

For $\psi=90$:

\begin{eqnarray}
F_{\sigma-\sigma^{\prime}}^{Mn}(0\frac{k}{2}0) & = & 2(f_{xx}^{(a)}-f_{yy}^{(a)}),
\label{eq: Lt str. factor 90 1}\\
F_{\sigma-\pi^{\prime}}^{Mn}(0\frac{k}{2}0) & = &
4f_{xy}^{(a)}\cos\theta_{B},
\label{eq: Lt str. factor 90 2}\\
F_{\sigma-\sigma^{\prime}}^{Mn}(0k0) & = &
2[(f_{xx}^{(a)}+f_{yy}^{(a)})-2f_{xx}^{(b)}],
\label{eq: Lt str. factor 90 3}\\ 
F_{\sigma-\pi^{\prime}}^{Mn}(0k0) & = &
4f_{xy}^{(b)}\cos\theta_{B}+2(f_{xz}^{(a)}-f_{xz}^{(b)})\sin\theta_{B}].\nonumber\\
\label{eq:LT str. factor 90 4}
\end{eqnarray} 

 For $\psi=0$:

\begin{eqnarray}
F_{\sigma-\sigma^{\prime}}^{Mn}(0\frac{k}{2}0) & = & 0,\,
F_{\sigma-\pi^{\prime}}^{Mn}(0\frac{k}{2}0) = 0,
\label{eq: Lt str. factor 0 2}\\
F_{\sigma-\sigma^{\prime}}^{Mn}(0k0) & = &
4(f_{zz}^{(a)}-f_{zz}^{(b)}),
\label{eq: Lt str. factor 0 3}\\
F_{\sigma-\pi^{\prime}}^{Mn}(0k0) & = &
4(f_{yz}^{(a)}-f_{yz}^{(b)})\cos\theta_{B}-2(f_{xz}^{(a)}-f_{xz}^{(b)})\sin\theta_{B}.\nonumber
\\
\label{eq:LT str. factor 0 4}
\end{eqnarray}

\subsubsection{On the Q+(01/20) reflections.}

As can be seen from equation \ref{eq: Lt str. factor 90 2}, only the
off-diagonal component $f_{xy}^{(a)}$ is measured in the
$\sigma-\pi^{\prime}$ channel at $\psi=90^{\circ}$. This is the same
matrix element that the $(0k0)_{\sigma-\pi^{\prime}}$ reflections are
sensitive to at \emph{high} temperatures (at low temperatures the
$(0k0)_{\sigma-\pi^{\prime}}$ reflections depend on the $f_{xy}$ term
of Mn$_{b}$, as the Mn$_{a}$ and Mn$_{c}$ $f_{xy}$ terms cancel with
each other). One can in principle therefore follow the temperature
dependence of the $f_{xy}^{(a)}$ term above and below the transition
by tracking the temperature dependence of the
$(0k0)_{\sigma-\pi^{\prime}}$ and
$(0\frac{k}{2}0)_{\sigma-\pi^{\prime}}$ peaks respectively.

Focusing on the energy lineshape, one observes a strong intensity for
$I(0\frac{3}{2}0)_{\sigma-\pi^{\prime}}$ at the absorption edge
(Fig. \ref{fig orbital spectra}). In particular, this energy
dependence is strongly reminiscent of the resonant signal observed
from the prototypical Jahn-Teller system $\textrm{LaMnO}_{3}$
\cite{Murakami98}, shown in Fig. \ref{fig orbital spectra lmo}. In the
context of the present model, the similarity between the
$\textrm{LaMnO}_{3}$ $(010)_{\sigma-\pi^{\prime}}$ and the
$\textrm{Pr}_{0.6}\textrm{Ca}_{0.4}\textrm{MnO}_{3}$
$(0\frac{3}{2}0)_{\sigma-\pi^{\prime}}$ spectra is explained naturally
by the fact that the Mn$_{a}$ and Mn$_{c}$ undergo a Jahn-Teller
distortion as a result of the orbital ordering while the Mn$_{b}$ and
Mn$_{d}$ maintain an in-plane square symmetry. At the
Q+$(0\frac{1}{2}0)$ reflections, the Mn$_{b}$ and Mn$_{d}$
contributions cancel because they are described with an in-plane
square symmetry. We note that this similarity of the energy lineshapes
is not a trivial result - the $(010)_{\sigma-\pi^{\prime}}$ lineshape
at high temperatures, which measures the same component of the tensor,
appears quite different: It has a smaller first resonance and a larger
second resonance (Fig. \ref{fig 010 RT} and \ref{fig orbital
spectra}).  Thus, the similarity of the resonances is a strong
evidence that the distortions - and thus the occupied orbitals - are
the same in the two cases. For $\textrm{LaMnO}_{3}$, the existence of
$3x^{\prime2}-r^{2}$/$3y^{\prime2}-r^{2}$ orbital order is
unquestioned and we conclude that it is the same orbital order
involved here on the Mn$_{a}$ and Mn$_{c}$ sites.

However, we do not believe the $3x^{\prime2}-r^{2}$ and
$3y^{\prime2}-r^{2}$-like orbitals are fully occupied. Indeed, as
discussed below, we will conclude, from an analysis of the Q+(010)
reflections that the charge disproportionation is incomplete, and that
there is a partial occupancy of the $x^{\prime2}-y^{\prime2}$ orbitals
on the Mn$_{b}$ and Mn$_{d}$ sites.

\subsubsection{On the Q+(010) reflections.}

For $(0k0)_{\sigma-\sigma^{\prime}}$, at $\psi=90^{\circ}$, one
measures the in-plane anisotropy, \emph{i.e.}
$f_{xx}^{(a)}+f_{yy}^{(a)}-2f_{xx}^{(b)}$ which is the difference
between the sum $f_{xx}^{(a)}+f_{xx}^{(c)}=f_{xx}^{(a)}+f_{yy}^{(a)}$
for the sites Mn$_{a}$ and Mn$_{c}$ and the same sum for the sites
Mn$_{b}$ and Mn$_{d}$, that is
$f_{xx}^{(b)}+f_{xx}^{(d)}=2f_{xx}^{(b)}$ (Eq. \ref{eq: Lt str. factor
90 3}). The observation of a resonance (see Fig. \ref{fig: (010) and
(030) DANES spectra}) indicates that
$f_{xx}^{(a)}+f_{yy}^{(a)}-2f_{xx}^{(b)}$ is non-zero. This implies
directly that the in-plane orbital occupancy is different on the two
sites.

An important question is whether this difference in the resonant
factors arises from a chemical shift of the \emph{1s} levels as
expected for a charge ordering. It is noteworthy that, as reported by
v. Zimmermann \emph{et al.}  \cite{Zimmerman01a} the resonant signal
from $F_{\sigma-\sigma^{\prime}}^{Mn}(0k0)$ disappears at
$\psi=0^{\circ}$. This indicates that $f_{zz}^{(a,c)}=f_{zz}^{(b,d)}$
(Eq. \ref{eq:LT str. factor 0 4}): thus there is no \emph{measurable}
difference in the out-of-plane configuration between the two sites.
As noted previously by Nakamura \emph{et al.}  \cite{Nakamura99} and
Garc\'ia \emph{et al.}  \cite{Garcia01} it is difficult to understand
the disappearance of the resonant signal at $\psi=0^{\circ}$, that is,
the equality of the out-of-plane resonant factors, $f_{zz}^{(a,c)}$
and $f_{zz}^{(b,d)}$, if the resonance involves a chemical shift of
the \emph{1s} levels, which would be expected to produce an isotropic
effect on the resonant factor. We discuss in the following the case
with and without a chemical shift of the \emph{1s} levels.

First, let us consider that there is a significant chemical shift of
the \emph{1s} levels. Then one would be forced to conclude that the
equality between $f_{zz}^{(a,c)}$ and $f_{zz}^{(b,d)}$ is accidental
and due to a relatively small magnitude of the \emph{1s} chemical
shift together with an out-of-plane population of the orbitals. This
would give rise to small and different displacements along the
\textbf{c} direction whose effect on the resonant factors would have
to exactly counterbalance the \emph{1s} chemical shift. For example,
the in-plane stretching of the $e_{g}$ $3x^{\prime2}-r^{2}$ -type
orbital on sites \emph{a} and \emph{c} brings the oxygens along the
\emph{z}-direction closer to Mn$_{a}$ and Mn$_{c}$ thereby raising the
resonance energy along the \emph{z}-direction. This brings the
spectrum of $f_{zz}^{(a,c)}$ closer to $f_{zz}^{(b,d)}$. In addition,
one could allow a population of the $3z^{\prime2}-r^{2}$ orbital on
the Mn$_{b,d}$ atoms (as predicted theoretically by van den Brink
\emph{et al.}  \cite{vdBrink99}). This would imply an increase of the
out-of-plane distance Mn-O along \textbf{c} on the sites \emph{b} and
\emph{d}, which would also lowers the \emph{4p} orbitals along the
\textbf{c} axis thereby shifting $f_{zz}^{(b,d)}$ closer to
$f_{zz}^{(a,c)}$.  However, this cancellation requires that the shift
of the \emph{1s} and \emph{4p} be precisely the same to cancel each
other.  Further, in $\textrm{Nd$_{0.5}$Sr$_{0.5}$MnO$_{3}$}$ one also
observes no resonance at $\psi=90^{\circ}$, thus requiring the same
accidental cancellation to occur in two different materials
\cite{Nakamura99}.  Thus, a chemical shift of the \emph{1s} level
appears unlikely.

Conversely, if one assumes there is no shift of the \emph{1s} level,
then the resonant signal at Q+(010) and Q+$(0\frac{1}{2}0)$ must arise
from structural distortions associated with the orbital
ordering. Indeed, as the spectra of the Q+$(0\frac{1}{2}0)$
reflections are similar to those of LaMnO$_{3}$, it seems likely that,
as discussed above, the signal at these reflections comes mostly from
the Jahn-Teller distortions, that is, an anisotropic energy shift of
the \emph{4p} levels. Furthermore, we show in the next section that
\emph{ab initio} calculations of the RXD spectra, for the distorted
structure, reproduce the lineshape of both the Q+$(0\frac{1}{2}0)$ and
Q+(010) reflections suggesting that they have a common origin (the two
types of reflections have different energy lineshape in part because
for the latter reflections there is interference with the non-resonant
Thomson scattering in the $\sigma-\sigma$ channel).

The different in-plane configuration explains the resonant signal when
the polarization points in the plane ($\psi=90^{\circ}$) and the
absence of any difference for the out-of-plane configuration explains
the absence of a resonant signal when $\psi=0^{\circ}$. It is likely
that the mean Mn-O distance is therefore the same along \textbf{c} for
all sites.

Finally, our results explain the azimuthal dependence of the intensity
at an $(0k0)_{\sigma-\sigma^{\prime}}$ reflection as arising mainly
due to the term $f_{xx}^{(a)}+f_{yy}^{(a)}-2f_{xx}^{(b)}$ (here we
have also used the expectation that the off-diagonal term
$f_{xz}^{(a,c)}-f_{xz}^{(b,d)}$ is negligible compared to the diagonal
terms). The intensity depends on the ratio of the resonant
contribution and the Thomson terms. At resonance, the signal is about
four times that of the Thomson scattering as deduced from the
off-resonant intensity. So the contribution from Thomson scattering
and resonant scattering of the Mn atoms are of a similar amplitude. In
this case, it is easily shown that the azimuthal dependence of the
$\sigma-\sigma$ component is $1+2\sin^{2}\psi+\sin^{4}\psi$ (again
ignoring the off-diagonal terms). Indeed, a two-fold dependence has
been measured for these reflections \cite{Zimmerman01a}.

High-resolution XANES studies have not shown a difference between the
scattering factors of the two different Mn sites at low temperature
\cite{Garcia01sg}. Conversely, as we have just shown, the RXD
measurements demonstrate that there are two differentiable Mn sites
with two different scattering factors. The limitation of the use of
the XANES technique for studies of small electronic reorganization
arises from the fact that it measures the sum of the contributions
from all sites. The different spectra are thus smeared out. This
limitation is worsened if the XANES measurements are performed on
powder samples because the spectrum is then the average over all
directions for which there are different absorption edges.  Even on
single crystals the differences in scattering power are difficult to
observe because of the presence of twinning. (We observed three
domains in our sample: an \textbf{a}-domain, a \textbf{b}-domain and
another domain propagating along the $(112)$ direction). Thus, XANES
measurements are inherently insensitive to small energy shifts of the
resonant factors. In contrast, RXD can directly measure the difference
in the scattering factors. Moreover the direction of the polarization
is well defined with respect to the crystallographic axes when
measuring a Bragg reflection of a specific domain. In the present
case, the sensitivity appears as the so-called ``derivative effect''
for the Q+(010) and Q+(100) reflections with the azimuthal angle
$\psi=90^{\circ}$, that is, the polarization is directed along the
\textbf{a} and \textbf{b} crystallographic directions respectively.
The XANES measurements do confirm that the chemical shift must be
smaller than 4.5 eV, the value obtained by comparing the parent
compounds. Such a large difference would be detected by XANES. Thus,
XANES does not support a complete charge disproportionation, in which
one assumes that the oxygen octahedra for the formal 3+ and 4+ Mn
atoms in Pr$_{0.6}$Ca$_{0.4}$MnO$_3$ are the same as in
$\textrm{La}\textrm{MnO}_{3}$ and $\textrm{Ca}\textrm{MnO}_{3}$
structures respectively \cite{Garcia01sg}. The XANES measurements are
however not inconsistent with a small disproportionation
\cite{Garcia01sg}.

To summarize, there is no evidence of a \emph{1s} chemical shift that
might arise from charge disproportionation, and the electronic
configuration may be regarded as an orbital ordering of the Mn on
inequivalent sites arranged in a checkerboard pattern.  The resonant
effect arises mainly due to a \emph{4p} shift induced by the
cooperative Mn-O distance modulations resulting from the orbital
ordering. We observe a Jahn-Teller distortion on half of the Mn
atoms. That is half the Mn atoms have an in-plane symmetry
($3x^2-r^2$-like) and half are symmetric in-plane. However, since
there is no strong evidence of charge disproportionation, we suggest
that the other half of the sites have a partially occupied $x^2-y^2$
orbital. To be more specific, one might write the electron
configuration of neighboring Mn atoms, A and B, as $|A;B>=\alpha|3d^4,
3z^2-r^2; 3d^3>+\beta|3d^3;3d^4,x^2-y^2>$, where $|3d^4,
3z^2-r^2;3d^3>$ refers to a configuration on Mn A with four electrons
in the \emph{d} band, the fourth electron in a $3z^2-r^2$-like
orbital, and with the neighbor Mn in a $3d^3$ configuration
\cite{Thomas03}. When $\alpha=\beta=\frac{1}{\sqrt{2}}$ there is
orbital ordering without charge ordering. Any occupancy of the
$3z^{\prime2}-r^{2}$ orbital must occur either equivalently on all
sites or with a random distribution, that is, the data exclude any
ordered difference along the \textbf{c} direction between the two
sites. Finally, we comment on the charge disproportionation. While
there is no direct evidence for significant disproportionation - in
the form of a chemical shift of the \emph{1s} levels - the charge on
the two Mn sites, as defined as the charge contained within a sphere
of given size, is unlikely to be the same on the JT distorted site as
on the undistorted site because of the varying bond lengths. Thus the
charge disproportionation is likely to be of the form
Mn$^{\nu-\delta}$ and Mn$^{\nu+\delta}$ where $\nu$ is the average
formal valence and $\delta < 0.5$. From the existing data set we
cannot set a lower limit on $\delta$.

In the next section we show that the \emph{ab initio} calculations of
the RXD spectra support the existence of a checkerboard pattern of
JT-distorted and regular oxygen octahedra.

\subsubsection{ab initio calculations
  with candidate crystallographic structures for half-doped
  manganites.}

\begin{figure}
\includegraphics[width=0.75\columnwidth]{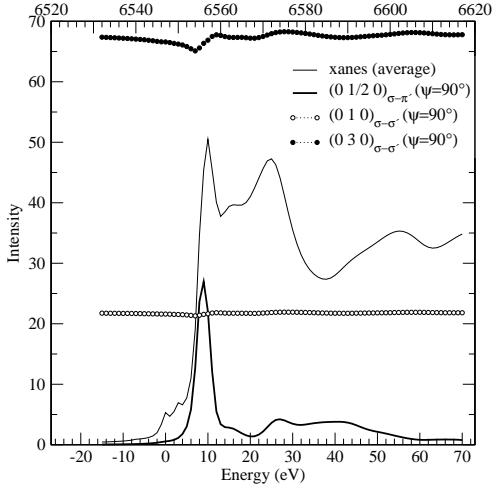}
\caption{\label{fig ab initio aziz}\emph{ab initio} calculations of
  the XANES, $(0\frac{1}{2}0)_{\sigma-\pi^\prime}$,
  $(010)_{\sigma-\sigma^\prime}$ and $(030)_{\sigma-\sigma^\prime}$
  spectra using the Pr$_{0.6}$Ca$_{0.4}$MnO$_3$ crystallographic
  structure from ref. \cite{Daoud-Aladine02}. The spectra are
  calculated with the same geometry than the data presented in figures
  \ref{fig: (010) and (030) DANES spectra} and \ref{fig orbital
  spectra}.}
\end{figure}

We performed the \emph{ab initio} calculations of the XANES and RXD
spectra in two crystallographic structures recently proposed for the
low temperature phase of the half-doped manganites, the so-called
``charge and orbital ordering'' phase. The FDMNES code and the same
procedure described in the section \ref{sec:Discussion HT} was
used. First, we used the refinement of Daoud-Aladine \emph{et al.}
\cite{Daoud-Aladine02} for Pr$_{0.6}$Ca$_{0.4}$MnO$_{3}$ which is
inconsistent with the checkerboard model. Second, we used the
structure of Radaelli \emph{et al.} for La$_{0.5}$Ca$_{0.5}$MnO$_3$
\cite{Radaelli97}, which reports a checkerboard model of the
inequivalent Mn atoms. We have not used the refinement of Lees
\emph{et al}. \cite{Lees98} for Pr$_{0.6}$Ca$_{0.4}$MnO$_{3}$ which
shows a checkerboard pattern of inequivalent Mn, because in this
proposed structure the $Q+(010)$ reflections have zero non-resonant
intensity in contradiction with the data.

The results for the structure proposed by Daoud-Aladine \emph{et al.}
\cite{Daoud-Aladine02}, the so-called Zener-polaron model are shown in
Fig. \ref{fig ab initio aziz}. For the $(010)_{\sigma-\sigma^\prime}$
and $(030)_{\sigma-\sigma^\prime}$ reflections, the resonant
contribution is at least one order of magnitude smaller than the
Thomson contributions of the other atoms, so the structure fails to
reproduce the strong resonant signal. We can understand the absence of
resonant effect for these calculations as follows. The proposed space
group is $P11m$ (monoclinic), with a strong orthorhombic $P2_{1}nm$
pseudosymmetry. The Mn$_{a}$ and Mn$_{b}$ atoms are then situated at
equivalent crystallographic sites and have the same valence and
orbital geometry. Instead, the inequivalent sites in the $P2_{1}nm$
space group are the ``A'' sites (in our notation, the Mn$_{a}$ and
Mn$_{b}$ sites) and the ``C'' sites (Mn$_{c}$ and Mn$_{d}$).  Such a
structure is inconsistent with the checkerboard-type model and is the
basis for introducing a Zener polaron model. The total structure
factor for the Mn atoms in this model is:

\begin{eqnarray}
F^{Mn}(h00) & \approx F^{Mn}(0k0)\approx &
-4(f_{xy}^{A}-f_{xy}^{C})\left(\begin{array}{ccc} 0 & 1 & 0\\ 1 & 0 &
0\\ 0 & 0 & 0\end{array}\right),\nonumber \\
\label{eq: P21nm str. factor}
\end{eqnarray}
with $h$ and $k$ odd. The approximation made here is that $e^{i2\pi
k\varepsilon}\approx1$, where $\varepsilon$ describes the small
relative displacement of the Mn atoms in the low-temperature unit
cell, $\varepsilon$ is typically of the order of $1/1000$
\cite{Daoud-Aladine02}. For the $F(h00)$ reflections, the diagonal
terms cancel exactly, and the model gives no resonant contribution in
the $\sigma-\sigma^{\prime}$ channel at any azimuthal orientation
(these reflections are indeed forbidden in the $P2_{1}nm$ space
group). For the $F(0k0)$ reflections, the diagonal terms don't cancel
exactly due to the small relative displacements between equivalent
atoms. However, the structure factor calculated from the structure
shows that the cancellation is still almost complete and the resonance
is damped.

Here, the empirical observation of a strong resonant effect for
$(0k0)$ reflections requires that Mn$_{a}$ and Mn$_{b}$ are in
different crystallographic sites and that they are surrounded by
different oxygen octahedra.  This deduction is independent of any
debate over the origin of the anisotropy, it is simply a statement
about the different anisotropies of the two sites and implies that
they can not be related by a mirror, a translation or a $\pi$
rotation. In the $P11m$ space group the positions of the atoms are
said to be almost the same as those in the $P2_{1}nm$ space group
\cite{Daoud-Aladine02}, that is equation \ref{eq: P21nm str. factor}
certainly still holds. Allowing significant differences in the atomic
positions would change equation \ref{eq: P21nm str. factor} but it
would also no longer represent the Zener polaron model. By going
toward a checkerboard model of inequivalent sites, one explains the
large spectroscopic effects observed experimentally.

\begin{figure}
\includegraphics[width=0.75\columnwidth]{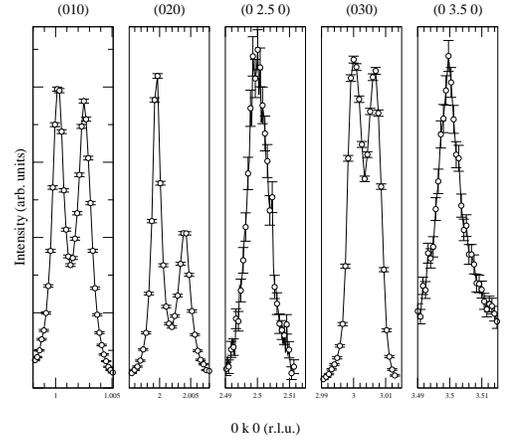}
\caption{\label{fig two domains} Evidence of $(k00)$ and $(0k0)$
  reflections, $k$ odd, in the low temperature phase (200K). The
  reflections were measured with a high-resolution Ge(111)
  analyzer. The splitting observed at the $(0k0)$ reflections is
  attributed to $(0k0)$ and $(k00)$ reflections due to the presence of
  a nearby twinned domain, the separation corresponds to two lattice
  parameters differing by 0.013\AA~ consistent with the difference in
  \textbf{a} and \textbf{b} lattice parameters at this temperature.
  At the $(0\frac{k}{2}0)$ reflections, there is no splitting as the
  doubling of the unit cell occurs only along the \textbf{b} axis. The
  data are not adjusted for differences in attenuation.}
\end{figure}

Also, as shown in Fig. \ref{fig two domains}, high resolution
measurements of the $(0k0)$ reflections reveal an apparent splitting
in reciprocal space. We attribute this splitting to the presence of
$(h00)$ and $(0k0)$ reflections from perpendicular (twin)
domains. However Q+(100) reflections are forbidden in the
$P2_{\textrm{1}}nm$ space group, and although the monoclinic $P11m$
permits the appearance of intensity at Q+(100) positions, in the
candidate structure they are expected to be much smaller than at
Q+(010) positions because of the pseudo-symmetry found with almost the
same positions \cite{Daoud-Aladine02}. In fact, we observe similar
intensities for these two types of reflections, together with a
similar, but not identical, energy dependence.  The observed splitting
corresponds to two lattice parameters, assumed to be \textbf{a} and
\textbf{b}, differing by $\Delta=0.013$ \AA.  This is consistent with
measurements of the in-plane lattice parameters (\emph{a} and
\emph{b}) in the COO phase: $a=5.4315$ \AA, $b=2\times5.4485$ \AA ~and
$c=7.6370$ \AA, so $\Delta=0.017$ \AA. Similarly Lees \emph{et al.}
reported $a=5.4313$ \AA, $b=2\times5.4413$ \AA~ and $c=7.6022$ \AA,
that is, $\Delta=0.01$ \AA~ at 200 K. So it seems very likely that
they are in fact \textbf{a} and \textbf{b} twin domains.

\begin{figure}
\includegraphics[width=0.75\columnwidth]{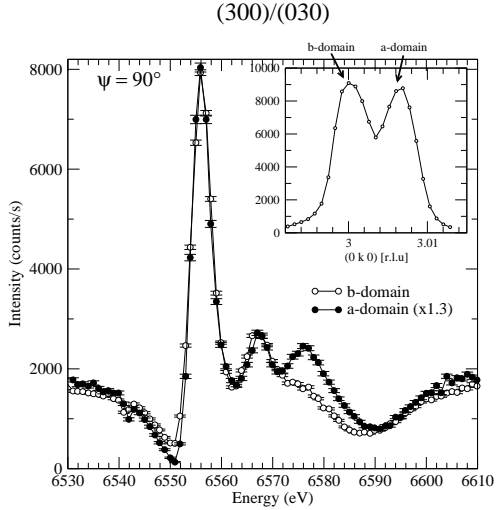}
\caption{\label{fig: 030 & 300}Incident energy dependence of the two
peaks observed at the (030) position at T=100K shown in the inset. The
peak at $k=3.00$ is attributed to the \textbf{b}-domains, \emph{i.e.}
to the (030) reflection, while the one at $k=3.006$ to the
\textbf{a}-domains, \emph{i.e.} to the (300) reflection. The two
spectra are scaled for clarity.}
\end{figure}

In Fig. \ref{fig: 030 & 300} we show the energy dependence for the two
peaks measured in the vicinity of $(030)$ at T=100 K. The $(300)$ and
$(030)$ reflections have similar intensity in the energy dependent
spectra which suggests that the displacements of the Mn atoms along
\textbf{a} and \textbf{b} are similar. The difference between the
spectra at 6576 eV presumably comes from the different polarization
directions along the \textbf{b} and \textbf{a} directions
respectively.  Interestingly, although the maximum is exactly at the
same energy position for both spectra (within the 1 eV/step
resolution), other features (like the pre-edge anomaly at 6541 eV)
seem to be shifted by 1 eV. In summary, it seems entirely plausible
that these peaks originate from different crystallographic domains and
that the actual structure gives equivalent Q+(100) and Q+(010)
reflections. We note that the reflections Q+(100) and Q+(010) are
present, with equivalent intensity, in the structure of the COO phase
of $\textrm{La$_{0.5}$Ca$_{0.5}$MnO$_{3}$}$ as refined by Radaelli
\emph{et al.} \cite{Radaelli97}.

\begin{figure}[t]
\includegraphics[width=0.75\columnwidth]{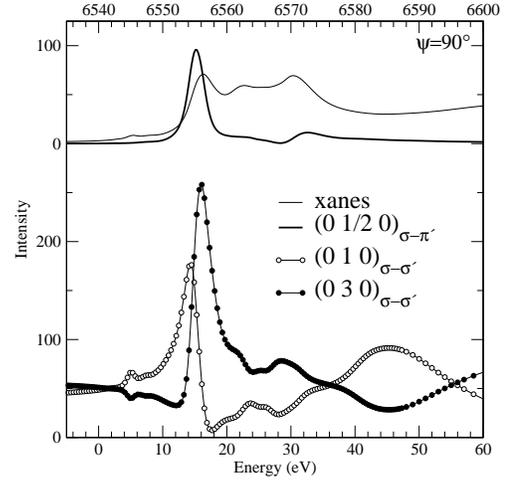}
\caption{\label{fig ab initio lcmo}\emph{ab initio} calculations of
  the XANES, $(0\frac{1}{2}0)_{\sigma-\pi^\prime}$,
  $(010)_{\sigma-\sigma^{\prime}}$ and
  $(030)_{\sigma-\sigma^{\prime}}$ with the structure of
  $\textrm{La}_{0.5}\textrm{Ca}_{0.5}\textrm{MnO}_{3}$, in the
  so-called charge and orbital ordering phase, as refined by
  Radaelli \emph{et al.} \cite{Radaelli97}. The results reproduce the
  data presented in figures \ref{fig: (010) and (030) DANES spectra}
  and \ref{fig orbital spectra}. The curves are rescaled for clarity.}
\end{figure}

\begin{figure}[t]
\includegraphics[width=0.75\columnwidth]{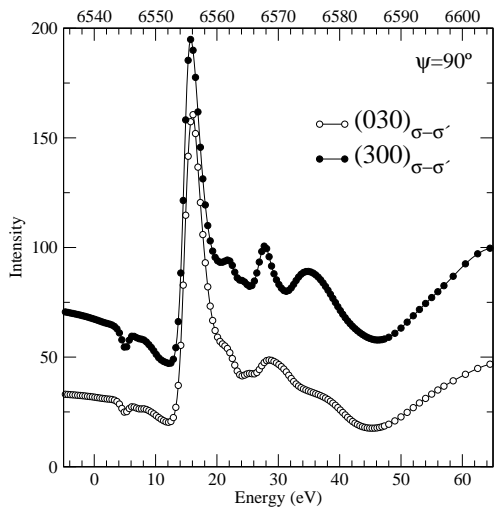}
\caption{\label{fig humps lcmo}\emph{ab initio} calculations of the
  $(300)_{\sigma-\sigma^{\prime}}$ and
  $(030)_{\sigma-\sigma^{\prime}}$ spectra with the structure of
  $\textrm{La}_{0.5}\textrm{Ca}_{0.5}\textrm{MnO}_{3}$
  \cite{Radaelli97}. The results reproduce the data presented in
  figure \ref{fig: 030 & 300}.}
\end{figure}

We show in Fig. \ref{fig ab initio lcmo} the results of the \emph{ab
initio} calculations with the crystallographic structure of
$\textrm{La}_{0.5}\textrm{Ca}_{0.5}\textrm{MnO}_{3}$ as refined by
Radaelli \emph{et al.} \cite{Radaelli97}.  The space group is
$P2_{1}/m$ and the inequivalent Mn are organized in the checkerboard
model, suggesting the CE-type ordering and consistent with the present
picture for $\textrm{Pr}_{0.6}\textrm{Ca}_{0.4}\textrm{MnO}_{3}$.  In
this structure, half of the Mn atoms have a Jahn-Teller distortion,
the other half are situated in regular, undistorted, octahedra. The
calculations of the XANES, the $(0\frac{1}{2}0)_{\sigma-\pi^\prime}$,
$(010)_{\sigma-\sigma^\prime}$ and $(030)_{\sigma-\sigma^\prime}$
reflection spectra are in very good agreement with the data
(Fig. \ref{fig: (010) and (030) DANES spectra} and \ref{fig orbital
spectra}). This structure also explains the observation of the
$Q+(100)$ and $Q+(010)$ reflections. In figure \ref{fig humps lcmo}
the calculated spectra for the (300) and (030) reflections are
shown. The fine structure, above the absorption edge, are in good
agreement with the data presented in figure \ref{fig: 030 & 300},
reproducing both the similarities and differences between the two
reflections between 6560 eV and 6590 eV. Also, the pre-edge region
below 6550 eV is well reproduced. The pre-edge feature marked by an
arrow in figure 5 is identified as a dipole transition because only
the dipolar operator is used in the calculations. However this feature
appears 9.5 eV below the absorption edge (maximum of the first
derivative) instead of 14 eV in the data. Presumably, this is in part
because of the approximation that the Mn atom is charge neutral,
$3d^54s^2$, thus details of the \emph{3d} band states will not be well
reproduced. Further calculations based on the finite difference method
beyond the muffin-tin approximation for the atomic potentials are
planed to improve the pre-edge spectra.  Such study is beyond the
scope of the present work.

These results suggest that the displacement of the atoms, in
particular the oxygen atoms, are equivalent in
Pr$_{0.6}$Ca$_{0.4}$MnO$_3$.  In particular, one notes that in the
La$_{0.5}$Ca$_{0.5}$MnO$_3$ structure the Mn-O distances along the c
direction are equivalent for all sites, explaining the absence of RXD
signal when the polarization of the photons is along the c
direction. In this
$\textrm{La}_{0.5}\textrm{Ca}_{0.5}\textrm{MnO}_{3}$ structure there
is no displacement of the Mn along the b direction, however a previous
report \cite{Zimmerman01a} on the observation of a resonant signal on
the $(0\frac{k}{2}0)_{\sigma-\sigma^{\prime}}$ reflections indicates
that the Mn atoms are actually also displaced along b at the phase
transition in Pr$_{0.6}$Ca$_{0.4}$MnO$_3$.

In conclusion, the \emph{ab initio} calculations of the RXD spectra
strongly support a checkerboard pattern of inequivalent Mn atoms, in
which half of the surrounding oxygen octahedra are Jahn-Teller
distorted and the other half have a nearly square in-plane
symmetry. Based on crystallographic and spectroscopic arguments, we
find that the structure invoked for the Zener polaron model can not be
correct. Instead, we determined that there must be a checkerboard
pattern of the inequivalent Mn atoms with one electron localized on
two Mn atoms.

\section{\label{sec:Summary.}Summary.}

We have attempted to determine the pattern and the local geometry of
the highest occupied orbital on the Mn sites in the near half-doped
manganite $\textrm{Pr}_{0.6}\textrm{Ca}_{0.4}\textrm{MnO}_{3}$ using
resonant x-ray scattering.  We have emphasized that resonant
diffraction can not be considered as a definitive probe if only
qualitative arguments are given, for example, the presence of a
resonant signal in the $\sigma-\pi$ channel and a particular azimuthal
dependence. Rather, a careful analysis of the resonant spectra is
required. This is especially true for the perovskite-type reflections
that mix charge, orbital and tilt orderings of the oxygen octahedra.

Based on such considerations, we have presented a model for the
low-temperature structure of Pr$_{0.6}$Ca$_{0.4}$MnO$_3$ which
describes an orbitally ordered structure of the CE-type. The CE
structure is stabilized by a slight structural distortion which arises
with the orbital ordering. Our experiments show that there is no
measurable chemical shift of the Mn \emph{1s} levels. However it is
likely that the orbital ordering implies a charge disproportionation
from the mean valence \emph{v} of the high temperature phase (as
defined as the charge contained around the two types of Mn
sites). Therefore a charge disproportionation
$\textrm{Mn}^{v-\delta}$and $\textrm{Mn}^{v+\delta}$ with $\delta <
0.5\,\textrm{e}^{-}$ might be considered. Unfortunately, at the
present stage of the analysis, the RXD spectra - measured at the Mn
K-edge - do not permit us to set a lower limit on $\delta$.

\section{Acknowledgments.}

We thank F. Bridges, A. Daoud-Aladine, J. Garc\'ia, J. Igarashi,
Y. Joly, J.E.  Lorenzo, P. G. Radaelli and T. Vogt for fruitful
discussions. We thank B. Ravel for the measurement of the absorption
spectra on X11A at the NSLS. Use of the Advanced Photon Source was
supported by the U.S.  Department of Energy, Office of Science, Office
of Basic Energy Sciences, under Contract
No.W-31-109-Eng-38. Brookhaven National Laboratory is supported under
DOE Contract No. DE-AC02-98CH10886. Support from the NSF MRSEC
program, Grant No. DMR-0080008, is also acknowledged.

\bibliographystyle{apsrev}
\bibliography{./pcmo_prb}

\end{document}